\documentclass[conference]{IEEEtran}
\ifCLASSINFOpdf
\else
\fi

\usepackage{amsfonts,amssymb}
\usepackage{amsmath}
\usepackage[ruled,vlined]{algorithm2e}

\SetCommentSty{mycommfont}
\usepackage{multicol}
\usepackage{multirow}
\usepackage{float}
\usepackage{graphicx}
\usepackage[dvipsnames]{xcolor}
\usepackage{url}
\usepackage{stfloats}
\usepackage{caption}
\usepackage{subcaption}
\begin{document}
%
\title{
Heterogeneous Dual-Core Overlay Processor for Light-Weight CNNs
}

\author{\IEEEauthorblockN{Tiandong Zhao, Yunxuan Yu, Kun Wang, Lei He}
\IEEEauthorblockA{Electrical and Computer Engineering Department\\
University of California, Los Angeles\\
}
}


%


\maketitle

\begin{abstract}
Light-weight convolutional neural networks (CNNs) have small complexity and are good candidates for low-power, high-throughput inference. Such networks are heterogeneous in terms of computation-to-communication (CTC) ratios and computation patterns between layers, especially for different layer types.
Yet, existing AI processors either use homogeneous processing elements (PEs), resulting in low runtime PE efficiency, or run different layers on heterogeneous PEs in sequential, introducing resource redundancy.
This paper proposes a heterogeneous dual-core architecture 
(dual-OPU), where one core is optimized for regular convolution layers and the other for depthwise convolution layers. PEs are homogeneous with each core. 
To make full use of dual-core parallelism, we develop a scheduling algorithm to concurrently execute layers for different input images on dual-core and balance parallel workload.
Meanwhile, we automatically tune the PE number for a core and tune the input size for each PE to maximize throughput. 
Compared with a single-core processor with the same area for a single network, heterogeneous dual-OPU on average improves runtime PE efficiency and throughput by 11$\%$ and 31$\%$, respectively.
For a workload of multiple networks, dual-OPU improves average throughput by 11$\%$ compared with the state-of-the-art processors scaled to the same area.
To the best of our knowledge, it is the first in-depth study on the heterogeneous dual-core processor for light-weight CNNs.

\end{abstract}
\IEEEpeerreviewmaketitle

\section{Introduction}
CNNs have achieved extensive success on miscellaneous artificial intelligence applications such as image classification and object detection.
A plethora of models emerge with different operators and architectures, gradually shifting attention from accuracy to efficiency in terms of speed and power. Classified as light-weight CNNs,
MobileNets \cite{howard2017mobilenets}\cite{sandler2018mobilenetv2} adopt depthwise separable convolution to reduce computation complexity and parameter amount, while other models, such as SqueezeNet \cite{iandola2016squeezenet}, alter model topology to spare computation power.

However, such model-level changes reduce runtime hardware efficiency.
To be more specific, layers in modern light-weight CNNs are heterogeneous concerning the computation pattern, especially between depthwise convolution layers and regular convolution layers.
Even for the same layer type,  CTC (Computation to Communication) ratio diffs significantly for various layer characteristic parameters such as input feature map size and kernel size.
Consequently, processors with uniform PEs (Processing Elements) can barely accommodate all types of layers efficiently, leading to a low runtime PE efficiency.
Separate engines for different layers, however, result in hardware resource redundancy due to the sequential execution of engines. 
In other words, the balance between generality and specificity has not been addressed well. This is particularly true
for workload with multiple CNNs.

There exist two paradigms on how to deal with the heterogeneous CNN workload.
One resorts to a uniform architecture, while the other applies custom architectures (often called accelerators) to different models.
For the first paradigm, GPGPU focuses on the acceleration of matrix multiplication with hardware multi-threading on parallel arithmetic cores and performs poorly on memory-bound depthwise convolution.
Similarly, TPU \cite{jouppi2017datacenter} is optimized to perform fast, bulky matrix multiplication with homogeneous systolic arrays.
Nevertheless, it suffers from poor performance on workload with sparse computation intensity, such as element-wise algebra.
Light-OPU \cite{yu2020light} utilizes uniform overlay architecture on FPGA for light-weight layers.
Yet, special-purpose modules, \textit{e.g.,} line buffer and squeeze-and-excitation block, might be redundant for the majority of layers, indicating resource inefficiency and potential room for further speedup.
Xilinx DPU \cite{dpudoc082019}\cite{wu2019high} also includes a non-negligible  additional resource cost for a separate depthwise convolution engine and channel augmentation module for small output channels.

On the other hand, accelerators like \cite{zhang2020dnnexplorer} adopt layer-wise  architecture 
tailored for bottom layers and uniform architecture for top layers based on the observation that CTC ratios fluctuate significantly between the two groups but just slightly within top layers.
\cite{wei2018tgpa}\cite{shen2017maximizing}\cite{li2016high} map one or some specific layers to heterogeneous convolution accelerators to handle various layer characteristics.
However, light-weight models are missing in these accelerators, so the performance is unknown when applied to latest light-weight models that are far more heterogeneous than traditional VGG-like models.
\cite{su2018redundancy} utilizes separate PEs for depthwise and regular convolution layers to improve individual throughput, which, in contrast, increases the overall redundancy in hardware resources.

To sum up, existing work with homogeneous PEs suffers from low resource efficiency on light-weight models, while those with heterogeneous PEs have not shown support for light-weight models or have exhibited resource redundancy.
To address these problems, we aim to find a balance between generality and specificity by proposing a dual-core processor (called dual-OPU), where one core is optimized for regular convolution and the other one for depthwise convolution with extra resource cost on light-weight hardware modules, \textit{e.g.,} line buffer.
Still, each core can handle all types of layers but with different efficiency.
To reduce the overall resource redundancy, we run multiple heterogeneous layers in parallel such that layers that prefer light-weight modules can be accommodated by a single core.
We interleave layers from different models to increase the chance of parallelizing heterogeneous layers.
Then we tune the PE numbers and PE input sizes of each core to balance the parallel workload, where the balance is further finetuned in tile granularity by layer split.
As a result, runtime efficiency is maximized.
Our contributions are listed as follows:
\begin{itemize}
    \item We propose a heterogeneous dual-core architecture with fine-grained PE configuration space for high run-time PE efficiency.
    \item We develop a scheduling algorithm to interleave layers from different models to exploit dual-core parallelism for high throughput.
    \item Given a set of target CNN models, we can find a PE configuration, with which high average throughput can be achieved.
\end{itemize}
\section{Motivation}
Light-weight CNN models typically include low-computation-intensity operators, \textit{e.g.,} depthwise convolution and pointwise convolution, to reduce operation count and parameter amount.
However, this model-level optimization leads to significant irregularity on the computation complexity between layers, where memory-intensive light-weight layers are interleaved with compute-intensive regular convolution layers.
Therefore, deployed on processor with uniform PEs, some layers have low runtime PE efficiency, as defined in Eq.\ref{eq:runtime-pe-efficiency}, where $N_{op}$ is the multiply-and-accumulate (MAC) operation amount performed in the measurement with frequency $f$.
$N_{PE}$ is the number of allocated PEs, and $\alpha$ is the number of MAC operations performed by each PE per clock cycle.
$T$ is the measured latency in seconds.
\begin{equation}\label{eq:runtime-pe-efficiency}
    \text{runtime\ PE\ efficiency} = \frac{N_{op}}{\alpha\cdot N_{PE}\cdot T\cdot f}
\end{equation}
Runtime PE efficiency measures the ratio between computation time and total latency of a layer.
Memory-intensive layers can barely fully overlap communication time with computation time, and thus have lower runtime PE efficiency than compute-intensive layers.
Furthermore, the gap between runtime PE efficiencies of different layers is exaggerated by variation of layer characteristic parameters such as input/output channels, input feature map height/width and kernel width/height.
\begin{figure}[!htb]
\resizebox{\columnwidth}{!}{
\includegraphics{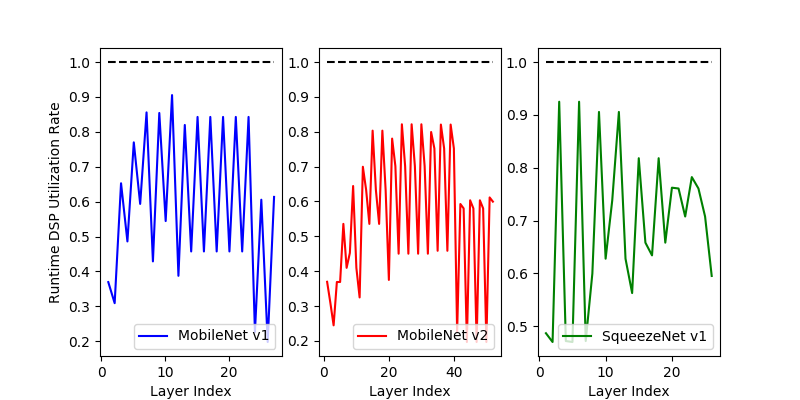}
}
\caption{Layer-wise runtime PE efficiency on uniform architecture proposed in \cite{yu2020light} with a single input image}
\label{fig:dsp_efficiency_v1_v2}
\end{figure}
Only when these layer-specific parameters match MAC unit sizes can we make full use of arithmetic resources.

To show the layer-wise runtime PE efficiency differences, we mimic the architecture of Light-OPU proposed in \cite{yu2020light} to measure the latency of each layer.
Industrial processors like TPU are not good candidates for layer-wise measurement due to their high runtime variance for an individual layer.
As shown in Fig.\ref{fig:dsp_efficiency_v1_v2}, the average runtime efficiency is 59$\%$, 41$\%$ and 62$\%$  for MobileNet v1, MobileNet v2 and SqueezeNet v1, respectively.
We find zigzag curves for all three models, where high efficiencies are contributed by regular convolutions and low ones come from depthwise convolutions for MobileNet v1/v2 and pointwise convolutions for SqueezeNet v1.
Both source layers of low PE efficiency have limited computing parallelism than regular convolution.
Devoid of output channel parallelism, depthwise convolution layers achieve 42$\%$ and 37$\%$ PE efficiency in MobileNet v1 and v2, respectively.
Pointwise convolution layers with small output channels lead to 41$\%$ PE efficiency in SqueezeNet v1.
The significant performance gap between different layer types calls for heterogeneous PEs, which customize arithmetic structures for specific layer types to improve runtime PE efficiency. 

Even with heterogeneous PEs, we cannot ignore that these light-weight models are almost purely sequential, which can hardly make full use of the parallel heterogeneous PEs on hardware.
With the emphasis on the throughput of multiple input images, prior work concurrently run layers for different input images for better resource efficiency.
\cite{wu2019high} runs MobileNet v2 in a layer pipeline schedule so that one pointwise convolution layer and one depthwise convolution layer for two input images run on different engines in parallel.
The fact that two layers run in parallel, however, still results in low PE runtime efficiency due to the imbalance of the latency even though its depthwise engine already utilizes small number of PEs to compensate.
Since its PE number for depthwise engine is not tunable, this imbalance worsens when the CTC ratios differ more significantly between two types of layers in MobileNet v1.
So we need to optimize PE allocation as well as layer scheduling.
\section{Heterogeneous Dual-core Architecture}
\subsection{Proposed Architecture}
To overcome the aforementioned problems, \textit{i.e.,} low runtime PE efficiency due to the heterogeneous workload and imbalanced schedule, we propose a novel dual-core architecture with fine-grained PE array configuration and an automatic design flow to find the best configuration for high throughput.
We define a core as computing unit with independent input/output buffers, a PE array and a post-processing unit. 
We introduce two types of cores,  channel-parallel core (c-core) and pixel-parallel core (p-core).
c-core exploits input/output channel parallelism for regular convolution, which usually has large channel numbers.
p-core takes advantage of the pixel parallelism in the kernel window for depthwise convolution, where line buffer is required as extra hardware support for data fetch due to the reuse of input feature map pixels from the sliding window.
PE configurations of the two cores are optimized with respect to layer characteristics for high runtime PE efficiency.
\begin{figure}[!htb]
\includegraphics[height=5cm,width=9cm]{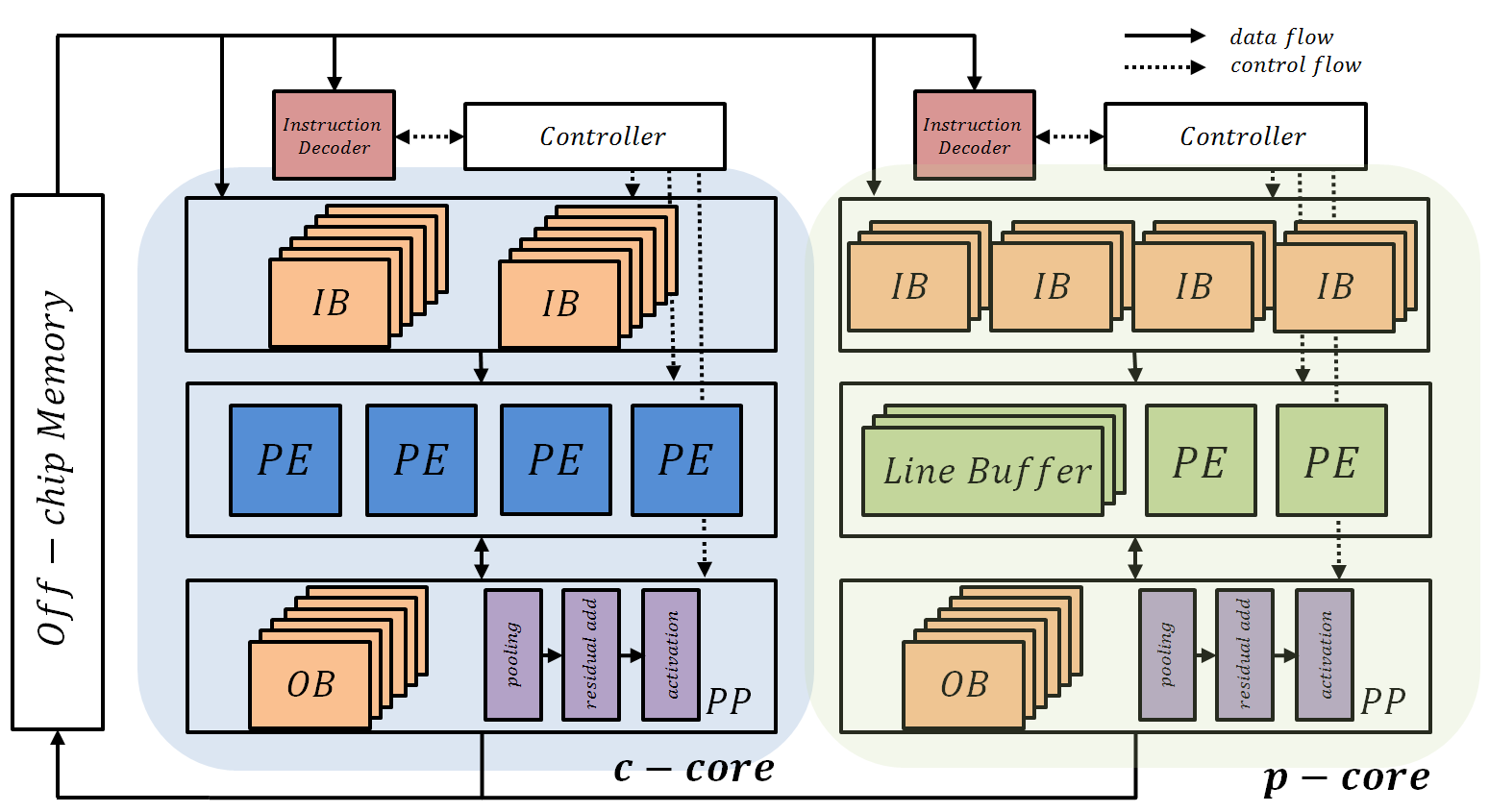}
\caption{Heterogeneous dual-core architecture. IB, OB and PP are input buffer, output buffer and post-processing unit.}
\label{fig:top-arch}
\end{figure}
As shown in Fig.\ref{fig:top-arch}, our heterogeneous architecture consists of one c-core and one p-core. Inside a core, PEs are homogenerous, and PE configuration and buffer sizes are customized.
Each core has ping-pong structured buffers for input feature maps, weights and biases.
Once a group of memory blocks for input feature map, weight and bias are loaded from off-chip memory to input buffers, the MAC operation pipeline is initiated.
Partial sums are stored in the output buffer for further accumulation.
Post-processing operations, such as pooling and activation, are also included in the computation pipeline and are initiated once a group of output feature maps is ready.

\subsection{PE Array Configuration}
We optimize PE array in terms of ($N_{PE}$, $N_{vector}$), where $N_{PE}$ is the number of PEs and $N_{vector}$ is the number of multipliers for each PE.
Each PE implements an inner product with $2\times N_{vector}$ inputs including $N_{vector}$ multiplication products, which are then reduced to one sum with a balanced adder tree.
More adders follow the PE outputs to provide 2 to $N_{PE}$ accumulated results.
The number of accumulated results is dynamically configured by instructions to obtain high runtime PE efficiency.

Regarding data fetch for c-core, input feature maps are duplicated and broadcast to PEs to exploit channel parallelism.
p-core has an alternative way of data fetch, by using line buffer. 
$T_{k_h}$ and $T_{k_w}$ are tiling sizes of kernel height and kernel width to compute for one single memory load from external memory to on-chip buffer. 
When $T_{k_h}$$>$$1$ or $T_{k_w}$$>$$1$, line buffer expands input feature maps by $T_{k_h}\times T_{k_w}$ before broadcasting.

Our PE may use multiple DSP macros on FPGA.  We decompose each DSP into two 8-bit multipliers to make full use of computation resources. 
However, the two decomposed multipliers must share of one input due to
hardware constraints.
Two multipliers in c-core share one input feature map pixel with two output channel weights, while two pixels share one input channel weight in p-core.
Double input feature map buffers are integrated with the p-core PE array.
As a result, two groups of sliding window pixels on the dimension of input feature map height are computed in parallel to make full use of DSP resources.

While c-core is more computationally powerful for channel parallelism,  p-core is flexible for both channel parallelism and pixel parallelism at the cost of some computation power and extra hardware components.
In addition to multipliers, p-core requires more resources for auxiliary components, such as line buffer and extra input feature map buffers, than c-core.
Processors using a single p-core not only leads to low runtime PE efficiency but also results in inefficient use of auxiliary components.
For example, when deploying regular $3\times3$ convolution with 64 input channels and 16 output channels on p-core P(16,64), we prefer not to use line buffer, since PE array configuration perfectly matches layer channel numbers.
Line buffer is not useful in such case.
If line buffer was used, it should generate multiple of $T_{k_h}\times T_{k_w}$ pixels as inputs of inner product, which ranges from 1 to 9 and leaves most of 64 PE inputs idle.
Therefore, we need to carefully allocate resource for p-core and c-core to make full use of all resources.

\subsection{Design Flow}
\begin{figure}[!htb]
\includegraphics[height=3cm, width=\columnwidth]{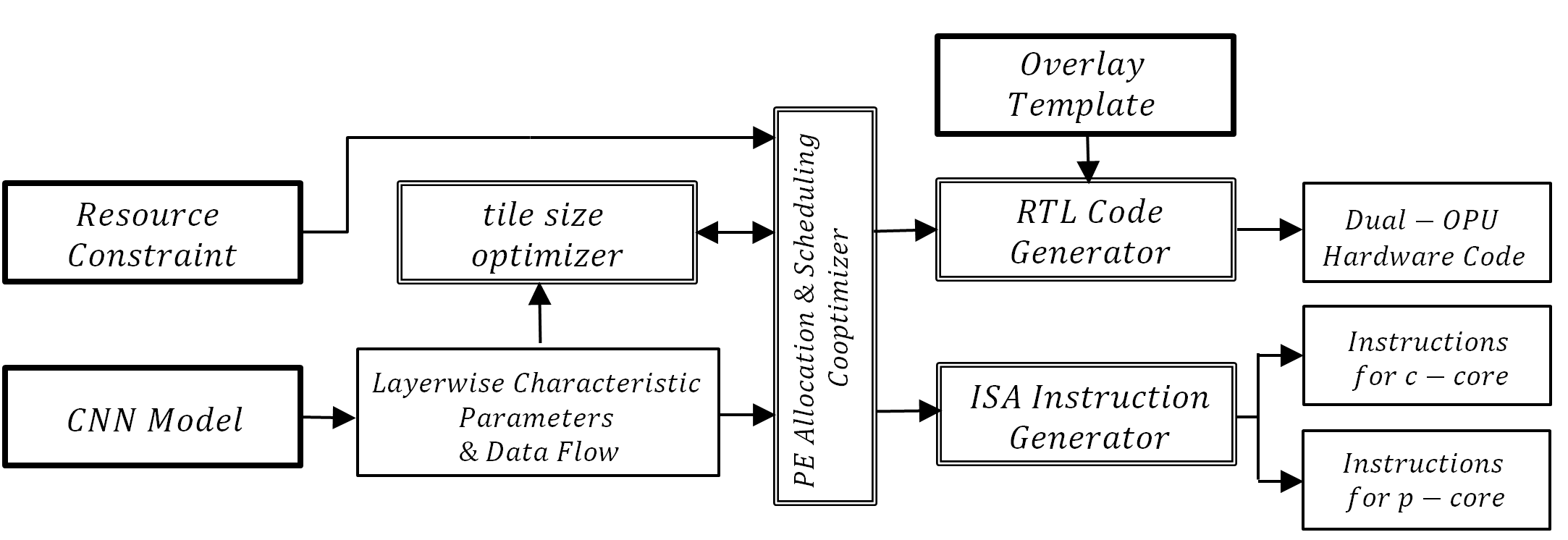}
\caption{Automation Flow}
\label{fig:workflow}
\end{figure}

To determine the resource allocation for c-core and p-core, we propose an automatic design flow to achieve high throughput of target workload.
Our flow in Fig.\ref{fig:workflow} takes CNN model description and FPGA resource budget as inputs.
CNN model description is parsed for layer-wise characteristics, such as input feature map height/width, input/output channels, kernel height/width, as well as data dependencies between layers.
We partition the layers into two groups, where layers in one group are assigned to c-core and layers in the other one are assigned to p-core.
For each group, we search for the PE configuration that leads to the highest runtime PE efficiency, under the constraint that the total allocated resources cannot exceed resource budget.
For each configuration, we decide tiling sizes on all dimensions with the objective to maximize runtime PE efficiency.
Based on the chosen tiling sizes, latency model and resource model are developed for c-core and p-core to estimate latency and resource.
Moreover, we balance the workload executed in parallel on c-core and p-core to improve the throughput.

One can see from runtime PE efficiency shown in Fig.\ref{fig:dsp_efficiency_v1_v2}, high efficiency and low efficiency are interleaved in all three cases. Adjacent layers with large difference are good candidates for potential balanced parallel execution.
To leverage this, we choose to interleave layers for two input images of the same CNN model and use topological order.
Then PE numbers on c-core and p-core are tuned based on the interleaved schedule to minimize the latency gap between parallel workload.
We further reduce the latency gap by a heuristic to split convolution layers along input feature map height dimension for tile reassignment.
Finally, we generate hardware code and instructions based on the PE allocation and scheduling.
\section{Modeling}
\subsection{Tile Sizing}
We aim to find tiling sizes of each layer for minimal total latency.
We use $(n, v)$ as the abbreviation of ($N_{PE}$, $N_{vector}$) in the following sections.
Given $(n, v)$, we aim to find the buffer sizes and if p-core is required for each layer so as to estimate the latency and the total resource.
Since tiling sizes compose the tensor size to load and compute at each time and are correlated with the memory and computation resource, we first determine the tiling sizes ($T_{c_i}$, $T_{c_o}$, $T_{k_h}$, $T_{k_w}$, $T_{h}$, $T_{w}$) for each layer.
Since each PE can implement inner product with multiple $v$ multiplications, product of $T_{c_i}$, $T_{c_o}$, $T_{k_h}$ and $T_{k_w}$ matches total multiplier count, as shown in Eq.\ref{eq:p-structure-multipliers}, where c-core has $T_{k_h}$=1 and $T_{k_w}$=1 without line buffer.
The multiple factor of $v$, denoted as $i$, aims to maximize runtime PE efficiency in Eq.\ref{eq:copy-mode}.
In other words, $i$ is the used number of PE among $n$ total PEs.
Larger $i$ indicates higher portion of PEs are utilized and thus higher runtime PE efficiency.
\vspace{-2em}
\begin{center}
\begin{equation}\label{eq:p-structure-multipliers}
    T_{k_h}\cdot T_{k_w}\cdot T_{c_i}\cdot T_{c_o} = n\cdot v,
    T_{k_h}\cdot T_{k_w}\cdot T_{c_i} = i\cdot v, i\in \mathbb{N}^+
\end{equation}
\begin{equation}\label{eq:copy-mode}
    i = \underset{i}{\arg\min} \lceil\frac{C_o}{T_{c_o}}\rceil\cdot  \lceil\frac{C_i\cdot K_h\cdot K_w}{T_{c_i}\cdot T_{k_h}\cdot T_{k_w}}\rceil
\end{equation}
\end{center}
We iterate $i$ from $1$ to $\lceil\frac{K_h\cdot K_w\cdot T_{c_i}}{v}\rceil$.
For each $i$, we iterate $T_{k_h}$ and $T_{k_w}$ to get $T_{c_i}$ = $i\cdot \lceil\frac{v}{T_{k_h}\cdot T_{k_w}}\rceil$ and $T_{c_o} = \frac{n\cdot v}{T_{k_h}\cdot T_{k_w}\cdot T_{c_i}}$.
($T_{h}$, $T_{w}$) relate to input buffer depth.
To simplify, we assume $T_h = T_w$, since most convolutions have square input size.
Eq.\ref{eq:buffer-depth-efficiency} decides ($T_{h}$, $T_{w}$) by memory efficiency in terms of buffer depth.
\begin{equation}\label{eq:buffer-depth-efficiency}
    (T_{h}, T_{w}) = \text{argmin} \frac{H\cdot W}{\lceil\frac{H}{T_{h}}\rceil\cdot \lceil\frac{W}{T_{w}}\rceil\cdot T_h \cdot T_w}
\end{equation}
It aims to minimize total input block numbers in size of ($T_h\times T_w$).
Some available options of tiling sizes might hold the same runtime PE efficiency, in which case we pick the one with less resource cost using the resource model to be discussed in Section \ref{section:resource-modeling}.

\subsection{Latency Modeling}\label{section:latency-modeling}
We calculate latency for each layer and then add them up. For each layer, $T_{load}$ and $T_{compute}$ are latencies of memory load and computation, respectively.
The three terms on the numerator of Eq.\ref{eq:load} are the data amount to load for input feature map in shape of ($H\times W\times C_i$), weight in shape of ($K_h\times K_w\times C_i\times C_o$) and bias in shape of ($C_o$), which are loaded sequentially through the limited bandwidth of external DRAM memory.
Memory load works in pipeline with $L_{dram}$ as the last part of latency.
$L_{dram}$ is the column address strobe (CAS) latency of DRAM access, and it stands for the delay between memory read request and the moment data is available for on-chip buffers.
Eq.\ref{eq:compute} depicts the computational latency as the product of tile numbers on each dimension. 
Computation following by post-processing runs in a deep pipeline.
Assuming that the data bitwidth is $BW_{data}$, whenever $min(C_o, \lfloor\frac{BW_{dram}}{BW_{data}}\rfloor)$ number of output feature map data are ready, they will be passed to post-processing unit and then stored to DRAM, resulting latency $L_{post}$.
Both $L_{dram}$ and $L_{post}$ are not constant in practice.
For estimation, we use average values based on multiple execution traces on FPGA. 
We use ceiling operators for the accuracy of modeling.
When the compiler generates and schedules ISA instructions, it aims to overlap $T_{load}$ and $T_{compute}$ as much as possible.
Therefore, we use the maximum between $T_{load}$ and $T_{compute}$ in Eq.\ref{eq:latency-total} to estimate latency of each layer.
\begin{equation}\label{eq:load}
    T_{load} = \lceil\frac{H\cdot W\cdot C_i+K_h\cdot K_w\cdot C_i\cdot C_o+C_o}{BW_{dram}}\rceil+L_{dram}
\end{equation}
\begin{equation}\label{eq:compute}
    T_{compute} = \lceil\frac{C_o}{T_{c_o}}\rceil\cdot \lceil\frac{H}{T_{h}}\rceil\cdot \lceil\frac{W}{T_{w}}\rceil\cdot \lceil\frac{C_i}{T_{c_i}}\rceil \cdot \lceil\frac{K_h}{T_{k_h}}\rceil \cdot \lceil\frac{K_w}{T_{k_w}}\rceil+L_{post}
\end{equation}
\begin{equation}\label{eq:latency-total}
    T_{total} = \sum\limits_{l}^{l\in layers} max(T_{compute}^l,T_{load}^l)
\end{equation}

\subsection{Area Modeling}\label{section:resource-modeling}
We will discuss the resource model that varies with PE configuration ($n$, $v$).
Total resource is the sum of the variants from PE and buffers and the invariant from memory controller, instruction decoder, post-processing unit, etc.
The variant cost on computation and memory resources are discussed as follows.
\paragraph{DSP}
We only use DSP to build multipliers for efficiency.
$\alpha$ is the MAC number that one DSP can handle.
In our design, each DSP48E1 slice is decomposed to two 8-bit$\times$8-bit multipliers that can work simultaneously with one common input.
So $\alpha$ is 2.
The required number of DSP is indicated as follows:
\begin{equation}\label{eq:dsp-modeling}
    N_{DSP} = \lceil\frac{n}{\alpha}\rceil\cdot v
\end{equation}

\paragraph{Memory Resource}
We assume PE array has two types of on-chip input buffers that are built by BRAM.
$B_{fm}$ is for feature maps and $B_{weight}$ is for weights.
Bias amount is usually small so it will be implemented by logic resource.
The buffer depth of $B_{fm}$ is $T_h\cdot T_w$.
Width of $B_{fm}$ and $B_{weight}$ are $T_{c_i}$ and $T_{c_i}\cdot T_{c_o}$ to match the input bandwidth of PE array.
For p-core, if depthwise convolution is applied, $B_{fm}$ should have 2 banks as that in the C-core model.
The depth of $B_{weight}$ is $\lceil\frac{C_o}{T_{c_o}}\rceil\cdot \lceil\frac{C_i}{T_{c_i}}\rceil$.
Xilinx FPGAs provide RAMB18K macro for 18kb block RAM.
Within total 18kb memory capacity, RAMB18K has configurable $width$ $\times$ $depth$ combinations, such as 36$\times$512, 18$\times$1k, 9$\times$2k, 4$\times$4k, 2$\times$8k and 1$\times$16k. 
We count the number of required RAMB18K given buffer width and depth with priority for width, which means that we tend to use minimum number of RAMB18K in term of width size.

\paragraph{Logic Resource}
LUT and FF cost comes from three aspects.
(1) the adders following multipliers in the PE array:
For each PE with $v$ inputs, they are accumulated to one with these adders.
Input data width of adder increases with the depth of the adder tree. 
(2) the delayers in a PE:
Delayers are required when $v$ is not power of two.
Delayers are implemented by simple register insertion.
(3) line buffer:
To match memory bandwidth with input buffer, $T_{c_i}$ is the number of line buffer channels required.
For each layer with $T_{k_h}\times T_{k_w} > 1$, the length of line buffer should be $T_w\times (T_{k_h}-1)+T_{k_w}$ such that pixels in sliding window $T_{k_h}\times T_{k_w}$ are preloaded before computation.
We select the parameter sizes to accommodate all the layers that require line buffers.
We collect resource costs from different sizes of adders, delayers and line buffers implemented by Xilinx toolchain and build model for each component.

\paragraph{Validation of Resource Model}
\begin{table}[!htb]
\centering
\resizebox{\columnwidth}{!}{%
\begin{tabular}{ c|cccc }
\hline
 & LUT & FF & DSP & BRAM\\\hline\hline
\cite{yu2020light} & 137816(67.29$\%$) & 251433(57.41$\%$) & 577(68.69$\%$) & 237.5(53.26$\%$)\\\hline
our resource model & 137149(67.62$\%$) & 234046(61.67$\%$) & 577(68.69$\%$) & 237(53.37$\%$)\\\hline
\end{tabular}%
}
\caption{Validation of resource modeling on core modules (external-memory-related modules excluded).}
\label{tab:resource-model-validation}
\end{table}
As shown in Table \ref{tab:resource-model-validation}, \cite{yu2020light} is a p-core with input/output buffers as well as other resource-invariant modules.
To show the effectiveness of our resource model on PEs and buffers, we compare the resource estimation with the real implementation results.
Our model is able to obtain $<$ $3\%$ resource estimation error.
\section{Optimization}
\subsection{Scheduling}
\begin{figure}[!htb]
\centering
\includegraphics[height=6cm,width=0.8\columnwidth]{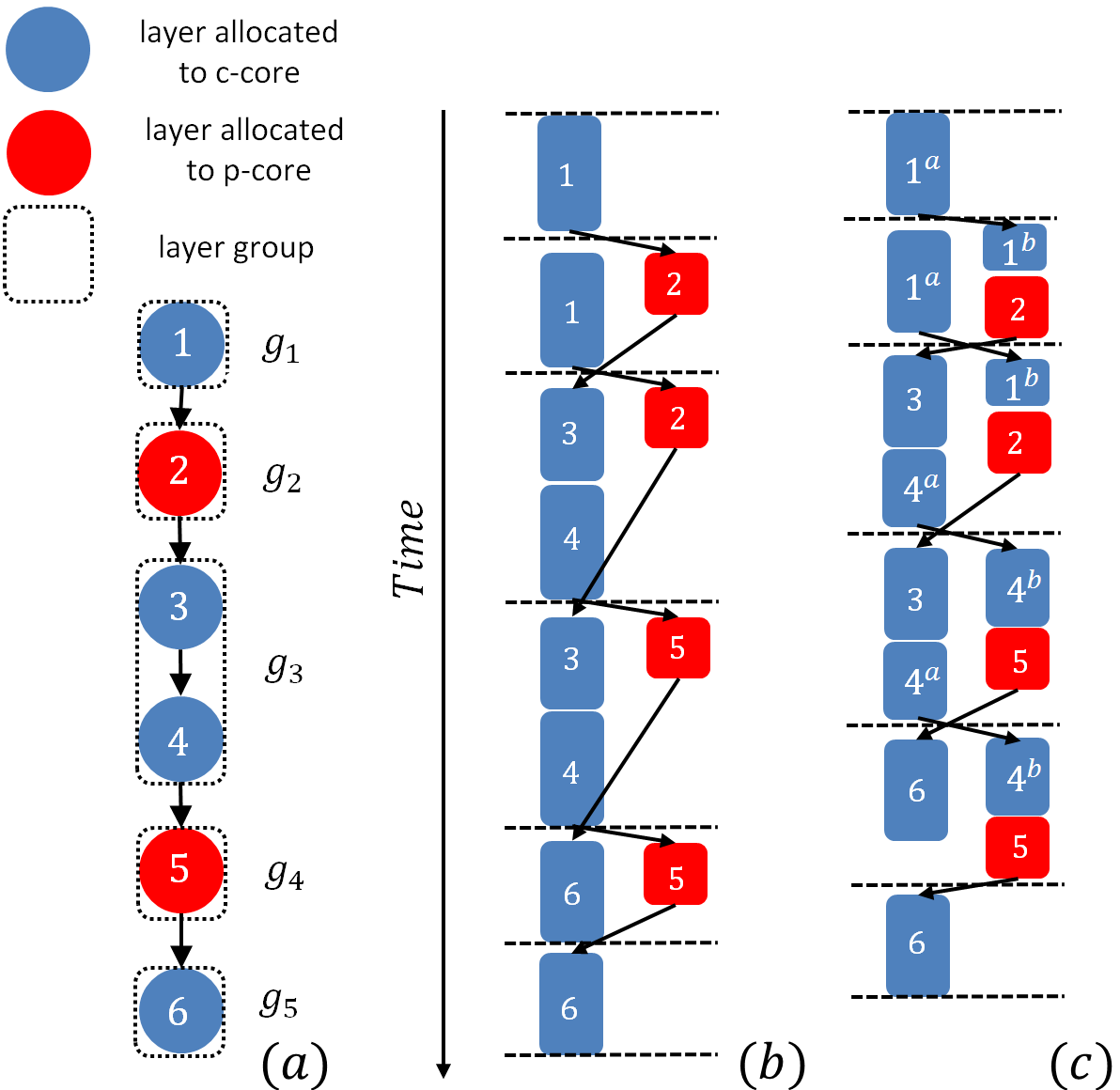}
\caption{Layer scheduling of MobileNet v2 snippet on heterogeneous dual-OPU (a) MobileNet v2 snippet where layer groups are assigned to c-core and p-core separately. (b) Execution trace where we interleave layer groups for two input images. (c) Execution trace where we further split layer 1 and layer 4 according to input feature map height to balance workload between two cores and improve runtime PE efficiency.}
\label{fig:scheduling}
\end{figure}

Given CNN graph $G(V,E)$ and separated ($n$,$v$) of c-core and p-core, we aim to find a schedule that maximizes the throughput of heterogeneous dual-OPU.
As shown in Fig.\ref{fig:scheduling}(a), nodes are layers, and edges indicate the data dependencies between layers.
We first partition the graph to layer groups, and each group includes one or multiple layers.
Groups are assigned to c-core or p-core following topological order.
Since the topology limits the chance for groups assigned to different cores to be executed in parallel (\textit{i.e.}, c-core-assigned $g_3$ and p-core-assigned $g_4$), we interleave layers for two input images of the same CNN graph such that more groups will be able to run simultaneously.
In Fig.\ref{fig:scheduling}(b), the second region from top indicates that $g_2$ for the first input image and $g_1$ for the second image are scheduled on p-core and c-core in parallel.
Due to the fact that layer characteristics and topology vary with different CNN graphs, two parallel groups can still show large latency difference even with the best allocation and partitioning scheme.
As a result, one core is idle during the gap, which lowers the performance.
We split some layers to sub-layers along the dimension of input feature map height to reduce the latency gap.
For example, layer 4 in Fig.\ref{fig:scheduling}(b) is split to layer $4^a$ and $4^b$ in Fig.\ref{fig:scheduling}(c).
Layer $4^a$ forms $g_3$ with layer 3, while layer $4^b$ forms $g_4$ with layer 5.
Consequently, latency gap between $g_2$ and $g_3$ is reduced.
Although latency gap between $g_3$ and $g_4$ increases in Fig.\ref{fig:scheduling}(c), split is applied as long as the total throughput is improved.
Among multiple partitioning choices with load balancing strategy, we pick the one with highest throughput estimation as the final schedule.

\subsubsection{Allocation-aware Partitioning}\label{section:layer-allocation}
We perform layer allocation followed by partitioning to find a scheme suitable for the scheduling method in Fig.\ref{fig:scheduling}(b).
We adopt three simple allocation schemes, including greedy allocation, layer-type based allocation and robin-round allocation.
Greedy allocation decides the core assignment based on the latency estimation of a layer on two cores.
Each layer is allocated to the core with less latency.
Regardless of hardware configuration, layer-type based allocation assigns regular convolution layers to c-core and depthwise convolution layers to p-core such that we can exploit channel parallelism and pixel parallelism, respectively.
As a work-around in case the two allocation scheme above cannot work, robin-round allocation assigns layers to two cores one by one in circular order.
We partition the graph to layer groups according to allocation results.
Interleaving layers for two input images, we aim to find such an allocation that the variance of $g_i/g_{i+1}$ for all odd $i$ and $g_{i+1}/g_i$ for all even $i$ is minimized.
Each group includes one or more layers instead of single layer for smaller variance.
Once we have a small variance, workload can be balanced well with appropriate PE allocation on c-core and p-core.
The partitioning result is shown in Fig.\ref{fig:scheduling}(a) as an example, where layer 3 and layer 4 form $g_3$.

\subsubsection{Load Balancing}
Partitioning in layer granularity cannot completely eliminate latency gap between all two parallel layer groups.
For example, a regular convolution layer is scheduled to c-core and run in parallel with a depthwise convolution layer with same parameter sizes.
The former one has far more MAC operations than the latter, leading to large latency gap when c-core and p-core have same MAC units.
To balance the workload, we propose a heuristic method to split layers.
Since input feature map pixels run in pipeline on either core, we pick the height dimension to split for simplicity, since it does not complicate the partial sum accumulation along input channels.
As shown in Alg.\ref{alg:load-balance-heuristic-scheduling}, we first compute latency gap for each parallel group pair and pick the pair with largest gap.
Once layer $l_{split}$ to split is located, we aim to find the height $h$ to remain in the layer to minimize the average latency of interleaved two batch runs $T_{b2}$, as defined in Eq.\ref{eq:b2-latency}.
$T_{b2}$ is the sum of the maximal latency between any parallel groups.
The input feature map size is updated from $H\times W$ to $h\times W$.
The rest of the layer is reallocated to the other core.
The height $h'$ accommodates the new $T_{k_h}$ in the reallocated part.
We continue splitting layers until there is no further improvement of $T_{b2}$.
\begin{equation}\label{eq:b2-latency}
    T_{b2} = \sum\limits_{i\in [1,N-1]}|T_{g_i}-T_{g_{i+1}}|+T_{g_1}+T_{g_N}
\end{equation}

\begin{algorithm}[!htb]
\SetAlgoLined
\KwIn{Layer groups $\{g_i\}$}
 \While{$T_{b2}$ gets improved}{
 compute $|T_{g_i}-T_{g_j}|$ for each neighboring ($g_i$,$g_j$)\\
 pick ($g_p$,$g_q$) with largest latency gap\\
 \tcc{assume $T_{g_p}$ $>$ $T_{g_q}$ and $g_p$ proceeds $g_q$} 
 find layer to split $l_{split}$ $\gets$ $g_p$.back()\\ 
 find height to remain $h$ $\gets$ $\underset{h\in [1,l_{split}.H-1]}{\arg\min} T_{b2}$\\
 $l_{split}.H \gets h$, $h' \gets l_{split}.H-h+l_{split}.T_{k_h}-1$\\ $g_q$.push$\_$front($l_{split}$.copy($H$=$h'$))\\
  update latency of each layer group and $T_{b2}$
 }
 \caption{Load-balance-heuristic Scheduling}
 \label{alg:load-balance-heuristic-scheduling}
\end{algorithm}

\subsection{Co-optimization of PE Allocation and Scheduling}
Aforementioned scheduling methods aim to obtain good throughput given arbitrary PE allocation of c-core and p-core.
Clearly, the PE allocation that better matches the heterogeneity of target workload leads to a higher throughput.
On the other hand, given PE allocation, we need to find a specific scheduling methods that can make full use the workload-matching PE allocation.
PE allocation and scheduling depend on each other.
We will first define the design space for PE allocation and then discuss how to find the best PE allocation along with scheduling for target workload.

\subsubsection{Design Space for PE Allocation}
Our proposed heterogeneous dual-core design is driven by PE array configurations of c-core and p-core. We pre-design memory buffers to meet bandwidth requirements of PE arrays, and
define parameter vector ($sch$,$n_c$,$v_c$,$n_p$,$v_p$) in Table \ref{tab:design-space-definition} as the design space 
for PE allocation, where
$sch$ specifies the scheduling with respect to layer allocation according to input model and hardware.
($n_c$,$v_c$) and ($n_p$,$v_p$) are PE configuration ($n$,$v$) for c-core and p-core, respectively.
For the constraint from target FPGA device, ($\overline{N}_{DSP}$,$\overline{N}_{BRAM}$,$\overline{N}_{LUT}$,$\overline{N}_{FF}$) stand for upper-bounds of DSP, BRAM, LUT, FF resources, which define the valid design space.
Another constraint is the bandwidth between core and external memory, $BW_{dram}$.

\begin{table}[!htb]
\centering
\resizebox{\columnwidth}{!}{%
\begin{tabular}{ |c|c|c| }
\hline
\multicolumn{2}{|c|}{Parameters} & Constraints\\
\hline
Scheduling & PE Allocation & ($\overline{N}_{DSP}$,$\overline{N}_{BRAM}$,$\overline{N}_{LUT}$,$\overline{N}_{FF}$)\\
\cline{1-2} 
$sch$ & ($n_c$,$v_c$,$n_p$,$v_p$) & $BW_{dram}$\\
\hline
\end{tabular}%
}
\caption{Design space for PE allocation. Tuning knobs are PE array configuration of c-core/p-core and scheduling.}
\label{tab:design-space-definition}
\end{table}

\subsubsection{Search Algorithm}
We use branch-and-bound to find the best PE allocation that minimizes the two-batch latency $T_{b2}$ within the resource constraints.
A naive approach is to determine if a DSP is included in our design along branches for all DSPs by enumeration, which, however, leads to huge search space in size of $2^{\overline{N_{DSP}}}$.
This search space is redundant in our dual-core design, since we only care about the DSP numbers allocated to c-core and p-core.
So, instead, we choose to branch upon the c-core DSP ratio $\theta$, as defined in Eq.\ref{eq:c-core-multiplier-ratio}, where
$\alpha$ is the MAC number one DSP Macro can perform per clock cycle.
\begin{equation}\label{eq:c-core-multiplier-ratio}
    \theta = \frac{n_c\cdot v_c}{\alpha\cdot \overline{N}_{DSP}}
\end{equation}
To compute the lower bound given $\theta$, we greedily allocate the remaining DSPs to p-core until we run out of logic resources or DSP resources.
We estimate the lower bound of $T_{b2}$ by Eq.\ref{eq:b2-latency}, where $T_{compute}$ is estimated with its lower bound of $T^{lb}_{compute}$ defined in Eq.\ref{eq:lower-bound-compute-time}.
$N^{core,l}_{DSP}$ is the DSP number of the core allocated for the layer $l$, which is $\theta\cdot \overline{N_{DSP}}$ for c-core and $(1-\theta)\cdot \overline{N_{DSP}}$ for p-core.
\begin{equation}\label{eq:lower-bound-compute-time}
    T^{lb}_{compute} = \frac{C_o\cdot H\cdot W\cdot C_i\cdot K_h\cdot K_w\cdot 2}{\alpha\cdot N^{core,l}_{DSP}}+L_{post}
\end{equation}
This is a lower bound of $T_{compute}$ because it does not include the potential unmatch between layer characteristic parameters and PE array configuration sizes, which results in higher latency.
We try different $sch$ based on the current $\theta$ and choose the lowest $T_{b2}$ as the lower bound for $\theta$. 
Then we branch to the two middle points of the unvisited subsets split by $\theta$.
We start with $\theta=0.5$ and search for the $\theta$ with minimal lower bound of two-batch latency.
The early termination happens whenever we reach the resource limit or we cannot have a better lower bound.

For the best $\theta$ found in the branch-and-bound global search, we then locally search for ($n_c$,$v_c$,$n_p$,$v_p$) for best throughput.
To reduce the search space, we select limited available options for $v$.
Although our PE array is able to provide runtime configurable $2$ to $n$ outputs, with fixed $n\times v$, small $v$ leads to huge cost on registers for accessibility of intermediate results.
On the other hand, as the unit input length, large $v$ can easily result in low runtime PE efficiency.
Therefore, we choose $\{8,9,10,12,14,15,16,18\}$ as candidates of $v$.
Prime numbers are excluded since common channel numbers are not multiple of prime numbers.
We exhaustively search all valid pairs of ($n$,$v$) for c-core and p-core based on the best $\theta$.
The ($n$,$v$) and $sch$ corresponding to the best throughput is our design choice.

\section{Experiment}\label{section:experiment}
\subsection{Experiment Settings}
\paragraph{Software}
We leverage parser of TVM \cite{chen2018tvm} framework that handles input models from different CNN developing frameworks (\textit{i.e.},  PyTorch, Tensorflow), and then transform Relay IR to our customized IR, which is used to generate ISA simar to that in \cite{yu2019opu} for dual-OPU.
\paragraph{Workload} 
Our test cases include MobileNet v1, MobileNet v2 and SqueezeNet v1.
We denote SqueezeNet v1 as SqueezeNet for simplicity.
These test cases cover typical light-weight operators such as depthwise convolution in MobileNet v1/v2 and expand/squeeze layers with small channels in SqueezeNet.
We set batch size as 2 for evaluation on throughput and
report on average values for each CNN model.
\paragraph{Hardware}
We use the notation C($n, v$) for c-core with a PE array with $n$ PEs, which have $v$ multipliers for each PE.
Same notation is applied to p-core, where PEs are further coupled with line buffers.
We run workload on three types of different designs, including single-core design P(128,9), homogeneous dual-core design P(64,9)+P(64,9) and heterogeneous design with one c-core and one p-core C(128,8)+P(64,9) to show the effectiveness of heterogeneous dual-OPU.
For fair comparison, PE configurations in each experiments below have same equivalent area.
To be more specific, three example designs have roughly same area, so are C(128,8) and P(64,9) in the heterogeneous one.
P(64,9) has half multipliers, buffer depth and line buffer channels of P(128,9).
C(128,8) has the same buffer depth as P(64,9).
Without line buffer, C(128,8) saves LUT resource for more multipliers than P(64,9).
Since line buffer only costs LUT while multipliers primarily cost DSP, it is difficult to compare the total resource cost.
To quantify the resource cost of PE array in c-core and p-core, we use the equivalent LUT cost as the equivalent area cost.
We count the equivalent LUT cost of each multiplier as the LUT cost to achieve the same functionality to one decomposed 8-bit multiplier implemented by DSP.
p-core PE array in P(64,9) and c-core PE array C(128,8) have close equivalent LUT cost as shown in Table \ref{tab:equivalent-lut-cost}.
P(64,9) has 128-channel line buffer to make full use of double input feature map buffers for extra pixel parallelism on the height dimension of depthwise convolution.
We use single p-core design P(128,9) as the baseline for comparison.
\begin{table}[H]
\centering
\resizebox{\columnwidth}{!}{%
\begin{tabular}{ |c|c|c|c|c| }
\hline
& \multicolumn{4}{|c|}{Equivalent Area/LUT Cost} \\\cline{2-5}
& Line Buffer & Multipliers & Adders & Total \\\hline
P(64,9) & 39868 & 40896 & 17859 &  98623\\\hline
C(128,8) & 0 & 72704 & 31749 & 104453 \\\hline
\end{tabular}%
}
\caption{Equivalent LUT cost of PE structures in P(64,9) and C(128,8). Similar total equivalent cost indicates similar area.}
\label{tab:equivalent-lut-cost}
\end{table}
\paragraph{Evaluation}\label{section:latency-validation}
We have built a cycle-accurate instruction level latency simulator with configurable core type (c-core/p-core), PE size ($n$,$v$) and memory bandwidth.
We run the complete compilation flow to generate ISA instructions for the simulation.
For each instruction, we adopts the latency model discussed in Section \ref{section:latency-modeling}, which takes CAS latency of DRAM access into account for accuracy.
As shown in Table \ref{tab:validation-trace-based-latency-simulator}, the cycle-accurate instruction-level simulator shows $<$1$\%$ error on cycle count compared with that on board-level FPGA implementation.
All the results in the experiment section are measured with cycle-accurate simulation.

\begin{table}[H]
\centering
\resizebox{\columnwidth}{!}{
\begin{tabular}{ |c|c|c| }
\hline
& \multicolumn{2}{|c|}{Cycle Count} \\
\cline{2-3}
& Board-level Performance & Cycle-accurate Simulator \\\hline
MobileNet v1  & 755857($\pm0\%$) & 757149(-0.2$\%$) \\\hline
MobileNet v2  & 637551($\pm0\%$) & 642940(+0.8$\%$) \\\hline
SqueezeNet  & 447457($\pm0\%$) & 443129(-0.9$\%$) \\\hline
\end{tabular}%
}
\caption{Validation of cycle-accurate instruction-level latency simulator on P(128,9).}
\label{tab:validation-trace-based-latency-simulator}
\end{table}

\subsection{Impact of Scheduling}
Table \ref{tab:scheduling-result} shows the effectiveness of our scheduling method on different ($N_{PE}$, $N_{vector}$) combinations.
C(128,8)+P(64,9) and C(180,8)+P(32,9) have different $N_{PE}$ ratio between two structures.
C(112,9)+P(72,8) further changes $N_{vector}$.
We compare the performance among four scheduling methods.
The first three only apply layer-type based allocation, greedy allocation and round-robin allocation for layer group partitioning.
Then we measure the average throughput of two interleaved batches.
The last one, load-balance-heuristic scheduling, further balances the parallel workload of layer groups on two cores based on the three aforementioned schedules.
We choose the best one as our final schedule.
Load-balance-heuristic scheduling improves throughput by 10$\%$ on average from the three basic schemes.
MobileNet v1 and v2 prefer layer-type based schedule as the basic scheme, while SqueezeNet gets more room for load balancing staring with round-robin scheme.

\begin{table}[!htb]
\centering
\resizebox{\columnwidth}{!}{%
\begin{tabular}{ |c|c|c|c|c|c| }
\hline
& \multirow{2}{*}{PE Array Configuration} & \multicolumn{4}{|c|}{Throughput(fps)}\\
\cline{3-6}
& & Layer-type & Greedy & Round-robin & Load-balance-heuristic \\\hline   
\multirow{3}{*}{MobileNet v1} & C(128,8)+P(64,9) & 267.4 & 267.4 & 269.8 & \textbf{304.3}\\
\cline{2-6}
& C(180,8)+P(32,9) & 318.9 & 259.3 & 266.6 & \textbf{320.2}\\
\cline{2-6}
 & C(112,9)+P(72,8) & 234.7 & 238.5 & 235.0 & \textbf{269.9}\\
 \hline
\multirow{3}{*}{MobileNet v2} & C(128,8)+P(64,9) & 378.4 & 378.4 & 338.5 & \textbf{427.6}\\
\cline{2-6}
& C(180,8)+P(32,9) & \textbf{392.0} & 304.9 & 214.4 & 384.9\\
\cline{2-6}
 & C(112,9)+P(72,8) & 323.7 & 346.6 & 317.0 & \textbf{371.1}\\
\hline
\multirow{3}{*}{SqueezeNet} & C(128,8),P(64,9) & 413.9 & 413.9 & 391.1 & \textbf{529.9}\\
\cline{2-6}
& C(180,8)+P(32,9) & 483.9 & 483.9 & 228.4 & \textbf{520.4}\\
\cline{2-6}
 & C(112,9)+P(72,8) & 328.3 & 375.2 & 372.5 & \textbf{451.3}\\
\hline
\end{tabular}%
}
\caption{Throughput comparison of different scheduling methods on different PE array configurations. Best throughputs are bolded.}
\label{tab:scheduling-result}
\end{table}

\subsection{Impact of PE Array Configuration}
Within the resource budget, Table \ref{tab:dse-result} shows the best PE array configuration for single workload.
\begin{table}[!htb]
\centering
\resizebox{\columnwidth}{!}{%
\begin{tabular}{ |c|l|c|l| }
\hline
& PE Array Configuration & DSP/PE Eff$^\dagger$ & Throughput/fps\\\hline
\multirow{2}{*}{MobileNet v1} & P(128,9) & 577/59$\%$ & 264.6($\pm0\%$) \\\cline{2-4}
& C(128,12)+P(8,16) & 832/$70\%$  &358.4($+35.4\%$) \\\hline
\multirow{2}{*}{MobileNet v2} & P(128,9) & 577/41$\%$ & 313.4($\pm0\%$)\\\cline{2-4}
& C(160,8)+P(48,8) & 832/$51\%$ & 438.4($+38.8\%$)\\\hline
\multirow{2}{*}{SqueezeNet} & P(128,9) & 577/62$\%$ & 446.9($\pm0\%$)  \\\cline{2-4}
 & C(130,8)+P(64,10) & 840/$75\%$ & 534.7($+19.6\%$)\\\hline
\end{tabular}%
}
\caption{Throughput comparison between PE array configurations optimized for single CNN and single-core baseline with same area. PE Eff$^\dagger$ is the runtime PE efficiency.}
\label{tab:dse-result}
\end{table}
We use single p-core design P(128,9) as the baseline.
Generated configurations have similar equivalent area in LUT cost on PE structure, including line buffer, multipliers and adders.
The results show that our design flow is able to generate PE array configuration with 31$\%$ improvement on throughput and 11$\%$ improvement on runtime PE efficiency on average over baseline.
On the generated configurations, we find the best configuration for MobileNet v1 holds the largest $\theta$,
indicating highest heterogeneity between parallel layer groups.
With largest workload difference between regular convolution and light-weight convolution, MobileNet v1 thus needs the largest $\theta$ among the three workloads to balance the load.
The result indicates that more heterogeneous workload can lead to more improvement by our heterogeneous dual-OPU.

\begin{table*}[!b]
\centering
\resizebox{2\columnwidth}{!}{%
\begin{tabular}{ |l|c|c|c|c| }
\hline
& \multicolumn{4}{|c|}{PE Array Configuration}\\
\cline{2-5}
& C(128,12)+P(8,16) & C(160,8)+P(48,8) & C(130,8)+P(64,10)  & C(128,10)+P(32,12)\\
& \multicolumn{3}{|c|}{Optimized for individual CNN} & Optimized for average throughput of multiple CNNs\\\hline
MobileNet v1 & \textbf{358.4}(+9.9$\%$)$^\dagger$ & 249.3(-23.6$\%$) & 314.6(-3.6$\%$) & 326.2($\pm0\%$)\\\hline
MobileNet v2 & 329.3(-24.8$\%$) & \textbf{438.4}(+0.2$\%$)$^\dagger$ & 428.1(-2.2$\%$) & 437.8($\pm0\%$)\\\hline
SqueezeNet & 527.9(+0.2$\%$) & 436.9(-17.0$\%$) & \textbf{534.7}(+1.5$\%$)$^\dagger$ & 526.6($\pm0\%$)\\\hline
Average$^*$ & 388.5(-6.1$\%$) & 349.6(-15.5$\%$) & 406.2(-1.9$\%$)  & \textbf{413.9}($\pm0\%$)$^\dagger$\\\hline
\end{tabular}%
}
\caption{Throughput comparison of PE array configurations on single-CNN workload and multiple-CNN workload. Average$^*$ stands for the average throughput of multiple-CNN workload. $\dagger$ indicates the configuration found by our search algorithm. Best throughput for each workload is bolded.}
\label{tab:dse-average-result}
\end{table*}
\begin{table*}[!b]
\centering
\resizebox{2\columnwidth}{!}{%
\begin{tabular}{ |l|c|c|c|c|c|c|c|c|c| }
\hline
CNN Model 
& \multicolumn{3}{|c|}{MobileNet v1} 
& \multicolumn{3}{|c|}{MobileNet v2} 
& \multicolumn{3}{|c|}{SqueezeNet} \\
\hline
Design
&\cite{bai2018cnn} &\cite{yu2020light} & Ours 
& Xilinx DPU \cite{dpudoc082019} & \cite{yu2020light} & Ours 
& Xilinx DPU \cite{dpudoc082019} & \cite{yu2020light}& Ours \\
\hline
Device 
& Stratix-V & XCK325T & XCK325T 
& ZCU102 & XCK325T & XCK325T 
& ZCU102 & XCK325T & XCK325T \\
\hline
PE Precision
& Int16 & Int8 & Int8
& Int8 & Int8 & Int8 
& Int8 & Int8 & Int8\\
\hline
Allocated DSP 
& 1278 & 704 & 832
& 2070 & 704 & 832
& 1942 & 704 & 832\\
\hline
Frequency(MHz)
& 133 & 200 & 200
& 287 & 200 & 200
& 333 & 200 & 200\\
\hline
Throughput(fps) 
& 237.1 & 264.6 & 326.2
& 587.2 & 325.7 & 437.8 
& 1048 & 420.9 & 526.6 \\
\hline
Throughput/DSP$^*$
& \multirow{2}{*}{0.11} & \multirow{2}{*}{0.21} & \multirow{2}{*}{0.23} 
& \multirow{2}{*}{0.08} & \multirow{2}{*}{0.14} & \multirow{2}{*}{0.16} 
& \multirow{2}{*}{0.20} & \multirow{2}{*}{0.19} & \multirow{2}{*}{0.22} \\
(GOPs/DSPslice)
& &  &  
& &  &  
& & &  \\
\hline
\end{tabular}%
}
\caption{Comparison between our work and existing works. Throughput/DSP$^*$ is normalized to 8-bit operations.}
\label{tab:compare-with-sota}
\end{table*}

Targeting on workload consisting of multiple CNN models, our flow is able to find the PE array configuration with higher average throughput than designs that are optimized for a single CNN model specifically.
We use harmonic mean of throughput for different models as the average throughput.
Our design flow finds C(128,10)+P(32,12) as the configuration with highest average throughput. 
As shown in Table \ref{tab:dse-average-result}, C(128,10)+P(32,12)  shows 7.8$\%$ improvement on average throughput of multiple CNNs with 3.8$\%$ performance loss on individual-best configurations for single CNN.
Results show the effectiveness of our design space exploration approach on searching for which configuration can lead to best average throughput.

\subsection{Comparison with state-of-the-art processors}
We compare our work with Xilinx DPUv3 \cite{dpudoc082019} and other processors from industry and academia, respectively, since they can handle regular convolution and depthwise convolution.
Xilinx DPUv3 is implemented on Xilinx ZCU102 board with three B4096$\_$EU cores.
In Table \ref{tab:compare-with-sota}, we include extra 48 DSPs per core in allocated DSP count for DPU on MobileNet v2 due to the depthwise convolution, besides the basic cost for regular convolution.
DPU performance on MobileNet v1 is not included since it has never been reported by Xilinx.
Scaled to same area, our heterogeneous dual-core processor improves throughput/DSP by up to 85$\%$ and 15$\%$ compared with the latest works from industry (Xilinx DPU) and academia.

\section{Conclusions and Discussions}
In this paper, we propose a heterogeneous dual-OPU to achieve high throughput of light-weight CNNs with high runtime PE efficiency.
In dual-OPU, one core is optimized for channel parallelism and regular convolution, and the other core is optimized for pixel parallelism and depthwise convolution.
Moreover, the PE number of each core and the input size of each PE can be tuned automatically with our design flow given target CNNs and FPGA device.
Meanwhile, we concurrently run layers for different input images of the same CNN and schedule with layer split to optimize the overall runtime PE efficiency.
The experiment shows that heterogeneous dual-OPU can improve throughput and runtime PE efficiency of homogeneous baseline with the same area by 31$\%$ and 11$\%$ for single CNN.
For a workload of multiple CNNs, compared with state-of-the-art processors, our design can improve the throughput by 11$\%$ on average than the latest works from industry and academia scaled to the same area.


\bibliographystyle{unsrt}
\bibliography{reference}

\end{document}